\documentclass[twocolumn,showpacs,preprintnumbers,amsmath,amssymb]{revtex4}

\usepackage{graphicx}
\usepackage{dcolumn}
\usepackage{bm}

\begin{document}

\preprint{}

\title{ 
Bagnold scaling, density plateau, and
kinetic theory analysis of 
dense granular flow
}

\author{Namiko Mitarai$^1$}\altaffiliation[Present address: ]{Department of Physics, Kyushu University.}
\author{Hiizu Nakanishi$^2$}
\affiliation{%
$^1$Frontier Research System,
RIKEN, 
Hirosawa 2-1, Wako-shi, Saitama 351-0198, Japan. \\
$^2$Department of Physics, Kyushu University 33, Fukuoka 812-8581, Japan.
}%

\date{\today}

\begin{abstract}
We investigate the bulk rheology of dense granular flow down a rough
slope, where the density profile has been found 
to show a plateau except for the
boundary layers in simulations [Silbert {\it et al.}, Phys. Rev. E
{\bf 64}, 051302 (2001)].  It is demonstrated that both the Bagnold
scaling and the framework of kinetic theory are applicable in the bulk,
which allows us to extract the constitutive relations from
simulation data.  The detailed comparison of our data with the
kinetic theory shows quantitative agreement for the normal and shear
stresses, but there exists slight discrepancy in the energy
dissipation, which causes rather large disagreement in the 
kinetic theory analysis of the flow. 
\end{abstract}

\maketitle

Flowing granular material behaves like a fluid,
but comprehensive understanding of its rheology is still far from complete.
In the low-density regime with large shear rate,
grains interact through instantaneous collisions
and are described by a hydrodynamic model 
based on kinetic theory of inelastic hard spheres \cite{rapid}.
As the system becomes denser,
the independent collision assumption
becomes questionable and the one particle distribution of
grain velocity may not be characterized by
a small number of parameters or temperatures.
When grains are nearly closed packed,
they may experience enduring contacts,
and the system behaves as in plastic deformation.

One of a few established laws that hold
for granular rheology up to the relatively dense regime
is the Bagnold scaling \cite{bag},
which states that the shear stress 
is proportional to the square of the strain rate.
In fact, this is the only possible form for the 
stress in the flow of rigid grains 
characterized by the 
shear rate $\dot \gamma$ and the packing fraction $\nu$,
because $\dot \gamma^{-1}$ is the only time scale.
Simple dimensional consideration gives the 
Bagnold scaling for the shear stress $S$ as 
\begin{equation}
S=A(\nu)m\sigma^{2-d}
\dot{\gamma}^2,
\label{eq:bagnold}
\end{equation}
where $m$ is the grain mass, $\sigma$ is the grain diameter,
and $d$ is the spatial dimension,
with $A$ being
a dimensionless coefficient 
that depends on $\nu$.
Obviously this scaling should have broad range of 
validity for a simple shear flow 
of cohesionless hard grains; 
it should hold until either the system 
becomes so dense that the elasticity of
particles comes into the problem or the 
shear banding destabilizes the uniform shear flow.
In the case of gravitational flow down a slope,
the shear is brought about by the gravity
and the gravitational acceleration $g$
brings another time scale into the problem,
but the Bagnold scaling is 
expected to hold in the denser region
where the effect of gravity on particle orbits between collisions
is not significant.
Actually the Bagnold scaling 
is observed 
in experiments \cite{bag,denseEXP}
and simulations \cite{denseDEM}
in the slope flows quite well.

\begin{figure}[t]
\includegraphics[width=0.47\textwidth]{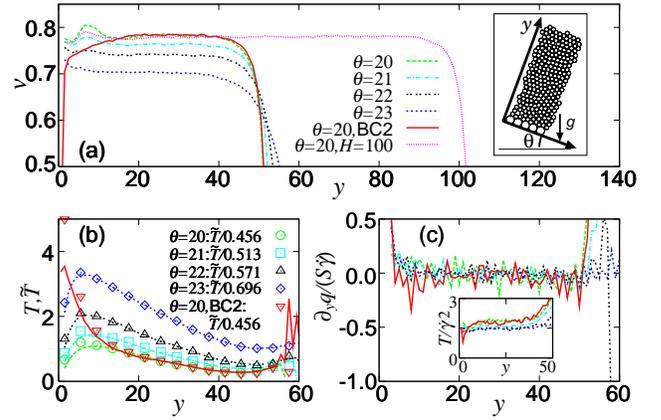} 
\caption{(color online)
The $y$ dependences of the packing fraction 
$\nu$ (a), the granular temperature $T$ (lines) 
and the rotational temperature $\tilde T$ (marks) (b),
and $\partial_yq/(S\dot \gamma)$ (c) for 
various inclination angles $\theta$.
The inset in (a) 
shows a schematic diagram of the system with a coordinate.
For most of the data, the 
bottom boundary is BC1 (see text)
and the total depth $H$ is 50,
but the data with BC2 and $H=100$ are also given for $\theta=20^\circ$.
In (b), $\tilde T$ are 
divided by the factors 
$0.456$ ($\theta=20^\circ$),
$0.513$ ($\theta=21^\circ$),
$0.571$ ($\theta=22^\circ$),
and $0.696$ ($\theta=23^\circ$)
to show $\tilde T \propto T $ in the bulk region.
The inset in (c) shows that $T/\dot\gamma^2$
 is roughly constant in the bulk (see text).
}
\label{profiles}
\end{figure}

Recently, Silbert {\it et al.} have performed 
large scale molecular dynamics simulations 
on dense slope flows \cite{denseDEM} and they 
found an interesting fact that
the grain density is almost constant and 
independent of the depth except for the boundary 
layers near the bottom and the surface.
This is quite intriguing because
the density is not nearly the closed packed density;
it depends upon the inclination angle $\theta$ but 
neither upon the total depth of the flow $H$ \cite{denseDEM} nor the 
roughness of the slope \cite{boundaryDEM}.
Somehow, the system adjusts its 
temperature to 
keep the density constant along the depth direction.

This is, however, not difficult to understand
if one extends the Bagnold scaling
to the pressure;
under the same condition,
the pressure, or
the normal stress $N$, should have the 
same form
\begin{equation}
N=B(\nu) m\sigma^{2-d}
\dot \gamma^2 
\label{eq:Ndimension}
\end{equation}
as $S$ with another dimensionless coefficient $B$,
thus the 
ratio of $S/N$ depends upon the packing fraction $\nu$,
but not on the shear rate $\dot \gamma$.
In the gravitational slope flow,
the force balance gives
$S/N=\tan\theta$, thus we have
\begin{equation}
A(\nu)/B(\nu)=\tan\theta,
\end{equation}
which shows the packing fraction is 
determined by the inclination 
$\theta$
but does not depend upon the depth $H$.
This argument suggests the existence of the
bulk region with the density plateau in the
gravitational flow is very general and
independent of detailed properties of grains.

In order to determine how the 
packing fraction $\nu$
depends on $\theta$,
we need a theory that gives 
constitutive relations.
This has been done by Louge \cite{L03} using
a kinetic theory for inelastic hard spheres 
\cite{Jenkins}. 
It is disappointing, however, to 
find that the kinetic theory fails to give
correct density profiles;
two branches of solution for $\nu$ were found,
but one gives too small $\nu$ and 
the other gives opposite $\theta$ dependence of $\nu$
to the one observed in simulations,
which implies the branch is a dynamically unstable one.
This is a little puzzling situation
because the kinetic theory has
been shown to hold in the case of sheared flow
in the similar density regime \cite{XuLouge}.

In this paper, we present detailed analysis of
our simulations on the bulk region of two-dimensional 
gravitational flow, assuming the framework of kinetic theory.
In contrast to previous works, where 
the overall profiles from hydrodynamic
models were discussed \cite{Profiles},
we examine each constitutive relations 
separately using data in the bulk
to avoid the uncertainty 
in a boundary condition for hydrodynamic 
equations.

First, we show how the bulk behavior is understood 
within the framework of kinetic theory.
In the granular kinetic theory,
the kinetic temperature 
$T \equiv m<({\bf c}- {\bf v })^2>/d$ is treated as 
a separate variable, which introduces an additional time scale.
Here, $\bf c$ is the particle velocity, 
$<\cdots>$ represents average over the microscopic scales,
and $\bf v=<\bf c>$.
The shear stress is given by the 
momentum flux; this is $m \ell(\nu) \dot\gamma n \sqrt{T/m}$
in the elementary transport theory,
where $n$ is the number density and $\ell(\nu)$ represents the mean free
path. More generally,
\begin{equation}
S=f_2(\nu)m^{1/2}\sigma^{1-d}
T^{1/2} \dot \gamma, \label{eq:s}
\end{equation}
with a dimensionless function $f_2(\nu)$,
which depends on $\nu$ and other 
material parameters such as a restitution coefficient.
Similarly  we have for the 
normal stress $N$, the energy dissipation $\Gamma$, 
and the heat flux $q$ 
\begin{eqnarray}
N&=&f_1(\nu)\sigma^{-d}T,\label{eq:n}\\
\Gamma&=&f_3(\nu) m^{-1/2}\sigma^{-d-1} T^{3/2},\label{eq:gamma}\\
q&=&-f_4(\nu) m^{-1/2}\sigma^{1-d}T^{1/2} \partial_y T,
\label{eq:heatflux}
\end{eqnarray}
with $\partial_y\equiv \partial/\partial y$.
The forms (\ref{eq:s}) to (\ref{eq:heatflux})
represent the quite general framework of kinetic theory \cite{rapid},
although functional forms $f_i(\nu)$ vary 
depending upon level of approximation.

These expressions should be compatible with
the Bagnold scaling when
the only relevant time scale is $\dot\gamma^{-1}$.
Equation~(\ref{eq:s}) of $S$ indicates 
that $T\propto \dot \gamma^2$ in order that the Bagnold
scaling (\ref{eq:bagnold}) should hold. 
In the kinetic theory,
$T$ is determined by the energy balance equation
\begin{equation}
-\partial_y q+S \dot \gamma-\Gamma=0 \label{eq:energy}
\end{equation}
in the steady flow. When the divergence of the heat flux
$-\partial_y q$ is zero as in the case of the uniform shear flow
due to the symmetry in the $y$-direction,
$T$ is determined by the local balance 
between 
the viscous heating $S\dot \gamma$
and the energy loss $\Gamma$; 
then the time scale that determines $T$ is the shear rate only. 
$S\dot\gamma=\Gamma$ with Eqs.~(\ref{eq:s}) and
(\ref{eq:gamma}) gives
\begin{equation}
T=\left[f_2(\nu)/f_3(\nu)\right] m\sigma^2
\dot \gamma^2,
\label{eq:kineticbag}
\end{equation}
thus the Bagnold scaling holds.
In the slope flow, the divergence of the heat flux
is not necessarily zero,
but it turns out to be small compared with 
the other terms.
Therefore, from Eqs.~(\ref{eq:s}), (\ref{eq:n}),
(\ref{eq:kineticbag}), and $S/N=\tan\theta$, we have
\begin{equation}
\tan \theta =\sqrt{f_2(\nu)f_3(\nu)}/f_1(\nu),
\label{eq:tannu}
\end{equation}
which gives $\nu$ as a function of $\theta$
if we know $f_i(\nu)$.

In the following, 
the above analysis is examined in detail
in comparison with the data of our 
two-dimensional simulations 
on the soft sphere model
with the disk mass $m$, 
the diameter $\sigma$, and the moment of inertia $I=m \sigma^2/10$
as in Ref. \cite{denseDEM}.
The particles stiffness is taken to be
in the region where the flow behavior
is already in the hard sphere limit
\cite{commentSoft}, which allows us to employ
the constitutive relations for hard disks
in the kinetic theory analysis in the following.
The linear spring-dashpot model and the Coulomb friction
with the coefficient $\mu=0.5$ are employed,
and the periodic boundary condition is imposed 
along the flow direction.
The bottom boundary is made rough
by attaching disks of diameter $2\sigma$, which we refer 
to as BC1: 
See Ref.\cite{denseDEM} for detailed descriptions of the 
model (Our model corresponds to the model ``L2'').
We confirmed that our data agree with theirs in the bulk,
although our slope length ($20\sigma$) is shorter
than theirs ($100\sigma$).
We show only the data for $\theta\ge 20^\circ$ with
$H=50$, which is well above the stopping angle $\theta_{\rm stop}$
($\theta_{\rm stop}\approx 18^\circ$ \cite{denseDEM});
the boundary effects become significant 
for $\theta$ closer to $\theta_{\rm stop}$\cite{boundaryDEM}.
The boundary effects are examined 
by simulating with
a slope covered with disks of diameter $\sigma$ (BC2).

To compare our data with the kinetic theory,
we use the normal restitution coefficient $e_p=0.92$ 
and the tangential restitution coefficient
$\beta=1$,
although the tangential restitution coefficient
in the simulation is not constant 
because of sliding collisions with 
the Coulomb friction \cite{collision}.
The Coulomb friction is important in simulation,
but no kinetic theories have been worked out yet with it in
two-dimension \cite{comment2D}.

\begin{table*}
\caption{\label{tab:consti}
The constitutive relations from kinetic theory in Ref. \cite{JR85},
with parameters
$\kappa=4I/(m\sigma^2)$,
$a=\kappa(1+\beta)/[2(1+\kappa)]$, 
$r=(1+e_p)/2$, and
$C=\left[
-(1-\tilde T/T)a^2+(5-8r)a+2(5-3r)\right]/2$.
$g_0(\nu)$ is the radial distribution function. 
}
\begin{ruledtabular}
\begin{tabular}{cc}
$f_1$&$(4/\pi)\nu(1+2r \nu g_0(\nu))$\\
$f_2$&$
\left(1/C g_0(\nu) \sqrt{\pi}\right)
(1+\nu g_0(\nu)(r+a))
\left[1+\nu g_0(\nu)
[(3r-2)r+
2ar-a^2(1+\tilde T/\kappa T)]
\right] 
+(4\nu^2 g_0(\nu)r/\pi^{3/2})
(1+a/2r)$ \\
$f_3$&$(4\nu^2g_0(\nu)r/\pi^{3/2})
\left[8(1-e_p)+
4\kappa(1+\beta)(1+e_p)^{-1}(1+\kappa)^{-2}
[2+\kappa (1-\beta)-(1+\beta)\tilde T/T
]\right]$
\end{tabular}
\end{ruledtabular}
\end{table*}

Figure~\ref{profiles} shows the 
$y$-dependence of $\nu$ (a),
$T$ (b), and $\partial_y q/(S\dot\gamma)$ (c) 
for various inclination angles $\theta$;
most of the data are for the system 
with the depth $H=50$ and BC1,
but the data for BC2 
and those for $H=100$ 
are also shown for $\theta=20^\circ$ 
for comparison.
The data are given in the unit system where the 
length $\sigma$, the mass $m$,
and the time $\sqrt{\sigma/g}$ are unities.
One can see that the packing fraction in the bulk does not depend on the 
depth, and the effects of the boundary condition are
confined within the boundary layer and the bulk properties are independent.
Figure~\ref{profiles}(c) shows that 
$\partial_y q$ is much smaller than
$S\dot \gamma$ in the bulk,
which is consistent with our argument to derive
Eq.~(\ref{eq:kineticbag});
the heat flux $q$ is estimated by 
the constitutive relation in Ref.\cite{JR85}.
The plots of $T/\dot \gamma^2$ in 
the inset shows that Eq.~(\ref{eq:kineticbag}) 
holds approximately. 

We compare our data with 
the constitutive relations 
derived by Jenkins and Richman \cite{JR85}
for two-dimensional inelastic hard disks.
The functions $f_1(\nu)$, $f_2(\nu)$, and $f_3(\nu)$ in the steady flow
are given in Table \ref{tab:consti}. 
We adopt the radial distribution function $g_0(\nu)$ 
from Ref. \cite{VTA03} 
\begin{equation}
g_0(\nu)=g_{c}(\nu)+\frac{g_{f}(\nu)-g_{c}(\nu)}
{1+\exp(-(\nu-\nu_0)/m_0)},
\label{eq:gl} 
\end{equation}
where
$g_{c}(\nu)= (1-7\nu/16)(1-\nu)^{-2}$ and
$g_{f}(\nu)=[(1+e_p)\nu(\sqrt{\nu_c/\nu}-1)]^{-1}$
with $\nu_c=0.82$, $\nu_0=0.7006$, and $m_0=0.0111$.

In $f_2(\nu)$ and $f_3(\nu)$,
the rotational temperature $\tilde T\equiv I<(w-\omega)^2>$ 
appears as $\tilde T/T$, where 
$w$ is the particle angular velocity and $\omega=<w>$:
In the kinetic theory, $\omega$ is simply assumed to be
$(\nabla \times {\bf v})_z/2$ \cite{JR85}, 
which holds except for the region near the bottom boundary \cite{MHN}. 
$\tilde T/T$ becomes constant in the kinetic theory \cite{JR85};
the value should be $1$ for our parameters, 
but is not in the simulations.
This should be mainly due to the Coulomb friction,
which has strong effects on particle rotations. 
In Fig.~\ref{profiles}(b), $\tilde T$'s 
are plotted by marks along with $T$'s (lines):
$\tilde T$'s are divided by
factors that give the best fits 
of $\tilde T$'s with $T$'s.
We see $\tilde T$'s fit with $T$'s in the bulk region 
with the factor around $0.5$, but the ratio depends on $\theta$.
In the following, we try both $\tilde T/T=1$ and 
the values obtained from the simulations
for $\tilde T/T$ in $f_2(\nu)$ and $f_3(\nu)$.

\begin{figure}[htb]
\includegraphics[width=0.3\textwidth]{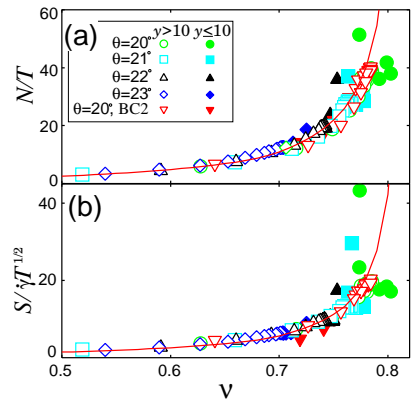} 
\caption{
(color online)
$N/T$ (a) and $S/(\sqrt{T}\dot{\gamma})$ (b) 
vs. $\nu$ for various $\theta$. 
The open and filled marks 
represent the data outside and within the bottom 
boundary, respectively.
The lines show
$f_1(\nu)$ (a)
and $f_2(\nu)$ (b). 
}
\label{f1f2}
\end{figure}

\begin{figure}[htb]
 \includegraphics[width=0.44\textwidth]{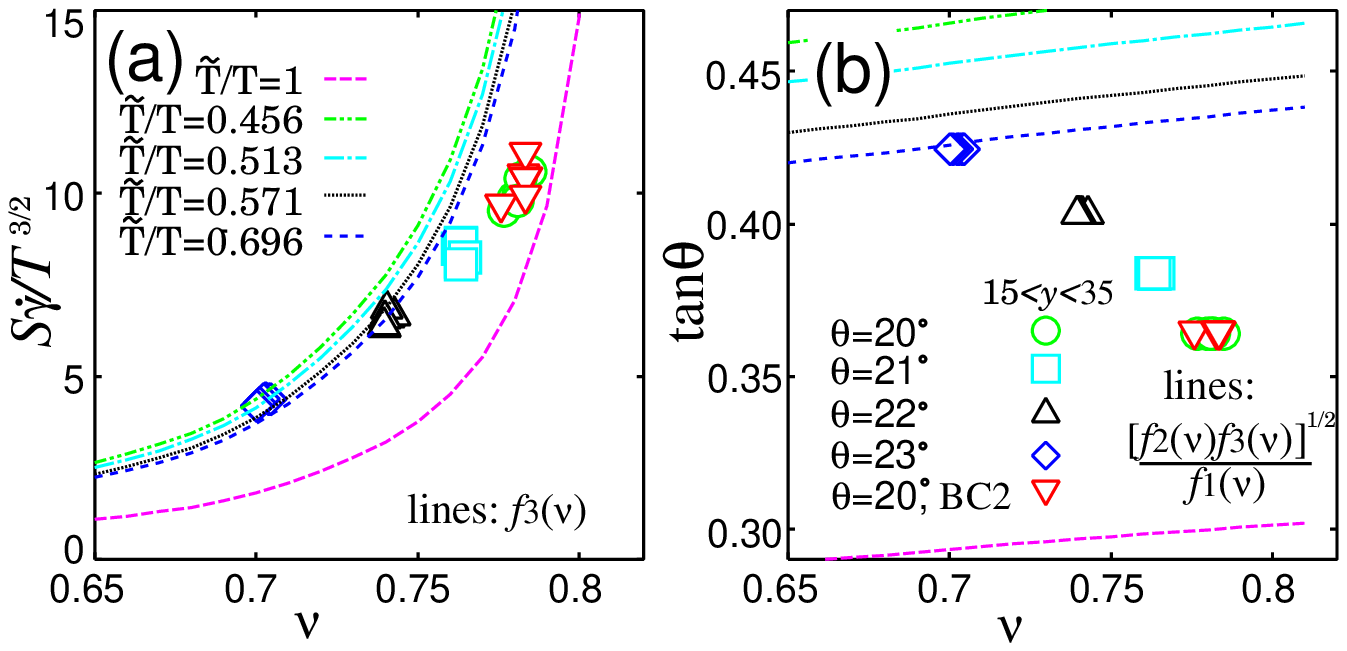} 
\caption{
(color online)
$S\dot{\gamma}/T^{3/2}$ vs. $\nu$ (marks) with
$f_3(\nu)$ (lines) (a), and $\tan\theta$
vs. the bulk density (marks) with 
$\sqrt{f_2(\nu)f_3(\nu)}/f_1(\nu)$ (lines) (b)
for various $\theta$.
The lines represent the plots with various values of
$\tilde T/T$. 
}
\label{f3}
\end{figure}

First, we examine Eqs.~(\ref{eq:s}) and (\ref{eq:n}).
Figure~\ref{f1f2} shows
the simulation data of 
$N/T$ (a) and $S/(\dot \gamma \sqrt{T})$ (b) against $\nu$ by marks.
The data from the bottom layers ($y\le 10$) are
distinguished by filled marks
because they follow a different trend.
The data outside the bottom layers 
($y>10$, open marks) for various $\theta$
collapse onto a single line.
This clearly shows that the expressions (\ref{eq:s}) and (\ref{eq:n})
in the kinetic theory are valid outside the bottom layers.
The data from the bottom layers show some scatter and 
different tendency between BC1 and BC2.

$f_1(\nu)$ and $f_2(\nu)$ in Table \ref{tab:consti}
with $\tilde T/T=1$ are shown by the solid lines 
in Fig.~\ref{f1f2}(a) and (b), 
respectively, and they agree with the data.
$f_2(\nu)$ depends on $\tilde T/T$ only weakly and 
the difference turned out to be
negligibly small in the range $0.5\lesssim \tilde T/T \lesssim 1$. 

Now, we examine $f_3(\nu)$ in Eq.~(\ref{eq:gamma}).
In Fig.~\ref{f3}(a), we plot
$S\dot{\gamma}/T^{3/2}$ against $\nu$ from the data;
this quantity should give $f_3(\nu)$ from
Eqs.~(\ref{eq:kineticbag}) and (\ref{eq:s}).
Only the data from the bulk ($15<y<35$) are plotted
because Eq.~(\ref{eq:kineticbag}) is valid only in the bulk
as we have already seen in Fig.~\ref{profiles}(c).
The lines show $f_3(\nu)$ from kinetic theory 
with various $\tilde T/T$.
$f_3(\nu)$ depends on $\tilde{T}/T$,
but the data agrees reasonably well 
with $f_3(\nu)$
when $\tilde T/T\sim 0.5$. 
Note, however, that the singularity at $\nu=\nu_c$
is weaker in the simulation data than in $f_3(\nu)$.

This difference in $f_3(\nu)$ is significant when we 
see them in the bulk density.
In Fig.~\ref{f3}(b), the
bulk density $\nu$ is plotted against the inclination angle $\theta$
for the simulation data and for the kinetic theory;
for the latter, we plot $\sqrt{f_2(\nu)f_3(\nu)}/f_1(\nu)$,
which should give $\tan\theta$ from Eq.~(\ref{eq:tannu}).
The bulk density decreases as $\theta$ increases in the simulation,
but $\sqrt{f_2(\nu)f_3(\nu)}/f_1(\nu)$ shows 
opposite tendency; the density increases as $\theta$ increases.
This discrepancy comes mainly from the discrepancy in $f_3(\nu)$,
more specifically from the fact that the data shows a weaker
divergence in $f_3(\nu)$, while the kinetic theory assumes
the same singularity in all of $f_1(\nu)$, $f_2(\nu)$, and $f_3(\nu)$
near the random closed packing. 

Some parts of the discrepancy might 
originate from the Coulomb friction, 
because it is included in the simulation
and should have some effects on energy dissipation,
but not taken into account in the 
existing two-dimensional theories.
The 
existing three-dimensional theory \cite{Jenkins},
however,
suggests that the way it changes $f_3(\nu)$
is just to modify
the coefficient of $\nu^2 g_0(\nu)$ as long 
as the level of approximation remains the same. 
Such a change
is not enough to make the singularity in $f_3(\nu)$ weaker.

It is a bit puzzling to find  
clear deviation from the 
kinetic theory in the energy dissipation while
the stresses agree quite well.
A possible origin of this is the
velocity correlation induced by the inelasticity,
which could violate the molecular chaos assumption 
in the kinetic theory; The decrease of the relative velocity 
tends to reduce the energy loss per collision.
This effect has been noticed in some
granular gas simulations, where the energy loss rate
is found to be more sensitive to the velocity correlation 
than stresses \cite{VelocityCorrelation}.
Careful analysis of the velocity 
correlation in dense flow is awaited.

Before concluding, let us make some comments on 
the Pouliquen's flow rule  \cite{denseEXP}:
The flow velocity at the surface scales
$u(H)/\sqrt{gH}=b H/H_{\rm stop}(\theta)$
with $H_{\rm stop}(\theta)$ being the depth of the flow below 
which the flow 
stops for a given inclination angle $\theta$,
and a numerical constant $b$ abound $0.136$. 
Erta\c{s} and Halsey \cite{Halsey} argued that the 
appearance of $H_{\rm stop}(\theta)$ in the expression of 
flow velocity for the depth $H,$
which can be much larger than $H_{\rm stop}(\theta)$, 
implies that the rheology of the 
dense gravitational flow is not local, and 
have proposed the eddy mechanism.
If the flow is controlled by a non-local mechanism, there 
is no way that the kinetic theory holds.
The Pouliquen flow rule, however, does not necessarily 
means a non-local mechanism but it simply means the stopping 
depth is determined by some aspects of the flowing rheology.
We do not know yet how it is determined,
but the present results suggest that 
the kinetic theory may well be a good starting point
to describe the flow.

In summary, by careful analysis of simulation data,
we have demonstrated that the rheology of gravitational 
dense granular flow
can be described within the framework of kinetic theory.
Especially, the constitutive relations based on  
the kinetic theory have been shown to 
agree quantitatively with the simulations,
but there is a slight discrepancy in the energy dissipation.
Due to this discrepancy, the kinetic theory fails to give
a correct description of the density plateau of 
the granular flow down a slope.

N.M. thanks M.Y. Louge 
for insightful discussions.
N.M. is supported by 
the Special Postdoctoral Researcher Program
in RIKEN. 
H.N. is supported by
a Grant-in-Aid for scientific research (C)
16540344 from JSPS.


\begin{thebibliography}{99}
\bibitem{comment2D}
Another problem about
the two-dimensional system might be
its strong tendency of crystallization.
In the shear flow of the present 
parameter region, 
however, the local structure is
changing in time and the system does not develop 
a long range order.
\end{thebibliography}
\end{document}